\documentclass[12pt]{article}
\usepackage{amsmath}
\usepackage{graphicx,psfrag,epsf}
\usepackage{enumerate}
\usepackage{natbib}
\usepackage{url} 

\usepackage{calc}
\usepackage{sidecap}
\usepackage{epsfig}

\usepackage{color}
\usepackage{amsgen}
\usepackage{amsbsy}
\usepackage{amsthm}
\usepackage{times}
\usepackage{ulem}
\usepackage{vmargin}
\usepackage[colorlinks,bookmarksopen,bookmarksnumbered,citecolor=black,urlcolor=black]{hyperref}
\usepackage{lineno}
\usepackage{setspace}
\usepackage{upgreek}

\usepackage{anyfontsize}
\usepackage{t1enc}
\usepackage[toc,page]{appendix}

\newcommand{\citeyearp}[1]{(\citeyear{#1})}
\newcommand{\bhat}[1]{\mathbf{\hat{\rm #1}}}
\newcommand{\pref}[1]{(\ref{#1})}

\newcommand{\btilde}[1]{\mathbf{\tilde{\rm #1}}}

\newcommand{\bc}{\ensuremath{\mathbf c}}

\newcommand{\br}{\ensuremath{\mathbf r}}

\newcommand{\bone}{\ensuremath{\mathbf 1}}
\newcommand{\bA}{\ensuremath{\mathbf A}}
\newcommand{\bB}{\ensuremath{\mathbf B}}
\newcommand{\bC}{\ensuremath{\mathbf C}}
\newcommand{\bD}{\ensuremath{\mathbf D}}

\newcommand{\bE}{\ensuremath{\mathbf E}}
\newcommand{\bF}{\ensuremath{\mathbf F}}

\newcommand{\bG}{\ensuremath{\mathbf G}}

\newcommand{\bI}{\ensuremath{\mathbf I}}

\newcommand{\bR}{\ensuremath{\mathbf R}}

\newcommand{\trace}[1]{\mbox{tr}(#1)}
\newcommand{\bU}{\ensuremath{\mathbf U}}

\newcommand{\bV}{\ensuremath{\mathbf V}}

\newcommand{\bX}{\ensuremath{\mathbf X}}
\newcommand{\bY}{\ensuremath{\mathbf Y}}

\newcommand{\bRxx}{{\bR}_{xx}}

\newcommand{\bRyy}{{\bR}_{yy}}

\newcommand{\bRxy}{{\bR}_{xy}}

\newcommand{\half}{\ensuremath{\frac{1}{2}}}

\newcommand{\beq}{\begin{equation}}
\newcommand{\enq}{\end{equation}}

\newcommand{\bali}{\begin{align*}}
\newcommand{\eali}{\end{align*}}

\newcommand{\bFs}{\bF_s}

\newcommand{\bGs}{\bG_s}

\newcommand{\bXs}{\bX_s}
\newcommand{\bYs}{\bY_s}

\newtheorem{theorem*}{Theorem}
\newtheorem{corollary*}{Corollary}[theorem*]

\begin{document}

\def\spacingset#1{\renewcommand{\baselinestretch}%
{#1}\small\normalsize} \spacingset{1}

  \title{\bf On the approximation of the between-set correlation matrix by canonical correlation analysis}
  \author{Jan Graffelman$^{\dagger,\ddagger}$ 1\thanks{
    \textit{This work was supported by the Spanish Ministry of Science and Innovation and the European Regional Development Fund under grant PID2021-125380OB-I00 (MCIN/AEI/FEDER). }}\hspace{.2cm}\\
  $^\dagger$Department of Statistics and Operations Research, Universitat Polit\`ecnica de Catalunya\\
  $^\ddagger$Department of Biostatistics, University of Washington\\
}
  \maketitle

\bigskip
\begin{abstract}
Canonical correlation analysis is a classic well-known multivariate statistical method focusing on the relationships between two sets of variables. The visualisation of those relationships can be achieved by means of a biplot of the between-set correlation matrix. The canonical analysis provides a low-rank approximation to the between-set correlation matrix that is optimal in a generalised least squares sense. This article proposes to adjust the between-set correlation matrix using either a single scalar effect, or column and/or row effects. An alternating generalised least squares algorithm is proposed to obtain optimal adjustments and low-rank factorisations. The adjustment leads to a better approximation of the between-set correlation matrix that achieves a lower root mean squared error in comparison with the classic canonical analysis. The results of the adjusted analysis can be efficiently visualised using biplots, with a minimal change in interpretation rules that only affects the biplot origin. Biplot calibration is used to enhance the visualisation of the results of the adjusted analysis. Some examples with publicly available data sets from social science, geochemistry and medical science illustrate the proposed improvement. Software for carrying out the adjusted canonical analysis in the R environment is provided.
\end{abstract}

\noindent%
\vfill

\newpage
\spacingset{1.45} 
\section{Introduction}
\label{sec:01}

Canonical correlation analysis (CCA) is a classic well-known multivariate method, due to Hotelling\citeyearp{Hotelling2,Hotelling3}, and designed to study the relationships between two sets of variables, with $p$ and $q$ variables respectively. Mathematically, the method identifies linear combinations for each of the two sets that have maximal correlation. CCA provides a generalised least squares (GLS) approximation to the between-set correlation matrix. The method is of profound theoretical and practical importance. Theoretically because CCA underlies several other multivariate methods that are concerned with two sets of variables, such as multiple and multivariate regression, correspondence analysis and linear discriminant analysis. Textbooks in multivariate analysis typically dedicate a chapter to the method~\citep{Johnson2,Mardia}. After its initial formulation in the thirties, the interpretation of the results of a canonical analysis has been greatly enhanced by the appearance of the biplot in the seventies~\citep{Gabriel}. CCA allows for a biplot the between-set correlation matrix, as proposed by~\cite{Haber}. These biplots were posteriorly developed in more detail by Ter Braak~\citeyearp{Braak13} and Graffelman~\citeyearp{Graffel16}. CCA has been generalised in multiple ways. Non-linear CCA~\citep{Burg} incorporated optimal scaling features in the canonical analysis. CCA has been generalised for multiple sets of variables~\citep{Carroll,Burg2}. Classic CCA inverts the two within-set correlation matrices, and consequently requires $n > \max(p,q)$, a restriction that has hampered its application in high-dimensional settings. For the high-dimensional setting, sparse canonical correlation analysis has been develop based on a penalised matrix decomposition~\citep{Witten1,Witten2}.
Gonzo\'{a}lez et al.~\citeyearp{Gonzalez} implement regularisation for two-set CCA. Takane~\citeyearp{Takane} incorporated regularisation into CCA with multiple sets. CCA has also been adapted to the compositional setting with the use of logratio transformations and generalised inverses~\citep{Graffel34}.\\

In many scientific studies there exists a split of the variables in one or more groups, and consequently CCA has found application in many areas of science. Classic examples are species counts and environmental variables for a set of sites in ecology~\citep{Gittins}; the relationship between genotypes (polymorphisms) and phenotypes (traits) in genetics~\citep{Osborne}. Other recent examples are measures of anxiety and the behavioural system in psychology~\citep{Wang}; the physicochemical traits and coagulation properties in milk in animal science~\citep{Caballero}; classic CCA and its extensions have been widely used in neuroscience~\citep{Zhuang} for the study of, for example brain activity and clinical measurements among many others.\\

The ordinary (within-set) correlation matrix is often approximated and visualised by means of biplot made by principal component analysis (PCA). In recent work, Graffelman and De Leeuw~\citeyearp{Graffel42} improved the approximation to the correlation matrix offered by PCA and other methods by using scalar adjustments and a weighted alternating least squares algorithm. Likewise, for the canonical analysis improvements in the approximation appear possible, for the between-set correlation matrix is neither centred nor adjusted in any other way prior to its factorisation by the singular value decomposition (SVD;~\cite{Eckart}), as evidenced by Eq.~\ref{eq:02} below.\\ 

The remainder of this article is organised as follows. We first develop some theory for adjusting rows and/or columns of the between-set correlation matrix and develop an alternating generalised least squares algorithm for carrying out the adjusted analysis. A subsequent section presents three different examples of an adjusted CCA. We provide software for the R environment~\citep{RRR} capable of doing the adjusted analysis. A discussion completes the article.

\section{Theory}
\label{sec:02}

A standard canonical correlation analysis can be formulated as the minimisation of the generic loss function

\begin{equation}
\upsigma(\bY) = \trace{\bR (\bX - \bY)\bC(\bX - \bY)'},
\end{equation}

where $\bX$ is the matrix of interest, $\bY$ its low-rank approximation and $\bR$ and $\bC$ are weight matrices. In the context of a canonical analysis, we have $\bX = \bRxy$, the $p \times q$ between-set correlation matrix, $\bR = \bRxx^{-1}$ the generalised least-squares weight matrix for the rows, $\bC = \bRyy^{-1}$ the generalised least-squares weight matrix for the columns, and $\bY$ the desired rank $k$ approximation to $\bRxy$. The solution of this minimisation problem is given by the singular value decomposition

\begin{equation}
  \bRxx^{-\half} \bRxy \bRyy^{-\half} = \btilde{\bA} \bD {\btilde{\bB}}',
  \label{eq:02}
\end{equation}

where $\bD$ contains the singular values in non-increasing order of magnitude, and $\btilde{\bA}$ and ${\btilde{\bB}}$ the orthonormal left and right singular vectors (${\btilde{\bA}}' \btilde{\bA} = \bI$ and  ${\btilde{\bB}}' \btilde{\bB} = \bI$). All main results of a canonical analysis follow from this SVD: the singular values in diagonal matrix $\bD$ equal the canonical correlations; matrices $\bA = \bRxx^{-\half} \btilde{\bA}$ and $\bB = \bRyy^{-\half} \btilde{\bB}$ contain the canonical weights for $X$ and $Y$ variables respectively; matrices $\bU = \bX_s \bA$ and $\bV = \bY_s \bB$ contain the canonical $X$ any $Y$ variates respectively, and where $\bXs$ and $\bYs$ contain the standardised variables of each set. Canonical biplots are constructed by jointly plotting the first two columns of

\begin{equation}
\bF = \bRxx \bA \bD_s^{\alpha} \quad \mbox{ and } \quad \bG = \bRyy \bB \bD_s^{1-\alpha},
\end{equation}

such that $\bRxy$ is factored as $\bRxy = \bF {\bG}'$, and where $\alpha$ ($0 \leq \alpha \leq 1$) is a conveniently chosen scalar. In practice, $\alpha = 1$ and $\alpha = 0$ are most often used, and the related coordinates are called standard an principal  coordinates respectively. For the given weighting, standard coordinates have unit sum-of-squares, whereas principal coordinates have a sum-of-squares that equals the squared singular values. The scaling of the coordinates may be conveniently indicated by a subscript (e.g.\, $\bF_s$ and $\bF_p$ for standard and principal coordinates respectively).

Equation~\pref{eq:02} shows that, despite the fact that the original {\it data} matrices in a canonical analysis are usually adjusted by subtracting means and dividing by standard deviations, the ensuing analysis of $\bRxy$ is essentially {\it unadjusted} as no column or row adjustments are applied to $\bRxy$. It immediately follows that the approximation to $\bRxy$ obtained in the classic analysis is, in general, sub-optimal. Geometrically speaking, Equation~\pref{eq:02} fits a hyperplane through the origin, whereas the cloud of correlations is typically not centred on the origin. A similar problem occurs in principal component analysis (PCA) which fits the correlation matrix sub-optimally with the spectral decomposition of $\bR$ as $\bR = \bV \bD \bV'$. The latter problem was recently resolved by Graffelman and de Leeuw~\citeyearp{Graffel40} by adjusting the correlation matrix with a scalar. In the canonical context, we suggest to improve the approximation of $\bRxy$ by adjusting its columns or rows with adjustment factors $\bc$ or $\br$, i.e., we propose to minimise 

\begin{equation}
\upsigma(\bY,\bc,\br) = \trace{\bR(\bX - \bY - \bone \bc' - \br \bone')\bC(\bX - \bY - \bone \bc - \br \bone')'}.
\end{equation}

For given $\bc$ and $\br$ the solution for $\bY$ is given by the SVD of

\begin{equation}
  \bR^{\half} (\bX - \bone \bc' -  \br \bone') \bC^{\half} = \btilde{\bA} \bD {\btilde{\bB}}'.
  \label{eq:05}
\end{equation}

After back-transforming the left and right singular vectors as $\bA = \bR^{-\half} \btilde{\bA}$ and $\bB = \bC^{-\half} \btilde{\bB}$, the rank $k$ approximation to $\bX$ is obtained
by

\begin{equation}
\bhat{\bY} =  \bA \bD {\bB}'.
\end{equation}

Setting partial derivatives $\partial \upsigma / \partial \bc$ and $\partial \upsigma / \partial \br$ to zero, the respective column and row adjustments are found to be:

\begin{equation}
  \bc = ((\bX - \bY)' \bR \bone  - \bone' \bR \br) /\bone' \bR \bone \quad \mbox{ and } \quad \br = ((\bX - \bY)  \bC \bone  - \bone' \bC \bc)/\bone' \bC \bone.
  \label{eq:07}
\end{equation}

We set up an alternating (generalised) least squares (AGLS) algorithm~\citep{DeLeeuw2022} for the adjusted canonical analysis that consists of the following steps:

\begin{enumerate}
\item Set convergence criterion $\epsilon$ to a small value, e.g. $1e^{-6}$.
\item Initialise $\bY^{(k)}$  and $\br^{(k)}$ (or $\bc^{(k)}$).
\item Using~\pref{eq:07}, calculate $\bc^{(k+1)} = ((\bX - \bY^{(k)})' \bR \bone  - \bone' \bR \br^{(k)}) /\bone' \bR \bone$. 
\item Using~\pref{eq:07}, calculate $\br^{(k+1)} = ((\bX - \bY^{(k)})  \bC \bone  - \bone' \bC \bc^{(k+1)})/\bone' \bC \bone$.
\item Use the SVD of~\pref{eq:05}, to calculate $\bY^{(k+1)}$.
\item If $\upsigma(\bY^{(k)},\bc^{(k)},\br^{(k)}) - \upsigma(\bY^{(k+1)},\bc^{(k+1)},\br^{(k+1)}) > \epsilon$ return to step (3) otherwise consider the algorithm has converged.
\end{enumerate}

This algorithm adjusts $\bRxy$ for both row and column effects. For practical purposes, it is better to adjust for column or row effects only, because a correlation matrix adjusted for column {\it and} row effects is, like a double-centred correlation matrix, particularly difficult to visualise because its biplot vectors do not have a unique origin~\citep{Graffel42}. This is easily achieved by setting $\bc$ or $\br$ to zero in the equations, and in that case steps (3) and (4) can be replaced by a single step updating column (or row) effects only.
Alternatively, one may also adjust $\bRxy$ for a single overall scalar effect ($\delta$) only. This has been developed and recommended for
a single set of variables by Graffelman and De Leeuw~\citeyearp{Graffel40} for it retains the symmetry of the low-rank approximation to the correlation matrix and it improves the fit in comparison with principal component analysis (PCA) and principal factor analysis (PFA). Because $\bRxy$ is generally neither square nor symmetric a single scalar adjustment is not compelling, though it may be used for simplicity if column or row adjustments do not give a substantial improvement of the fit. With a single scalar the loss function becomes

\begin{equation}
\upsigma(\bY,\delta) = \trace{\bR(\bX - \bY - \delta \bone \bone')\bC(\bX - \bY - \delta \bone \bone')'},
\end{equation}

and the scalar adjustment is given by

\begin{equation}
  \delta = \frac{\bone' \bR (\bX - \bY) \bC \bone}{\bone' \bC \bone \bone' \bR \bone}.
  \label{eq:09}
\end{equation}

The optimal value for $\delta$ can be found with minimal modifications of the algorithm given above, using this new loss function, adjusting $\bX$ in Eq.~\pref{eq:05} with $\delta \bone \bone'$, and replacing steps 3 and 4 using the Equation~\pref{eq:09} for updating $\delta^{(k+1)}$.

After convergence, it is desirable to quantify the amount of error of the low-rank approximation that has been fitted by the algorithm. For this purpose one can either report the value of the loss function or the root-mean-squared error (RMSE) of the approximation. Taking into account the weighting by generalised least squares, the RMSE is given by

\begin{equation}
  \mbox{RMSE} =  \sqrt{\trace{\bR \bE \bC {\bE}'}/(pq)} = \sqrt{\upsigma/(pq)},
  \label{eq:10}
\end{equation}

where $\bE = \bRxy - \bhat{\bR}_{xy}$ contains the errors of the approximation. The RMSE is just the square root of loss divided by the number of entries in the between-set correlation matrix. The RMSE has the advantage that it quantifies the amount of error in the correlation scale. With $\bR = \bI_p$ and $\bC = \bI_q$, Eq.~\pref{eq:09} gives the RMSE according to an ordinary least squares (OLS) criterion; consequently, when the canonical analysis is applied to the principal components of the $X$ and the $Y$ variables, the RMSE obtained by GLS and OLS will be the same.\\   

In a biplot obtained by CCA, a pair of orthogonal vectors between an $X$ and a $Y$ variables implies, assuming a perfect fit, that these variables are uncorrelated; i.e., the origin represents correlation zero. When a single scalar adjustment $\delta$ is used, the two orthogonal vectors represent variables that have correlation $\delta$. When column adjustments $\bc$ are used, the origin represents correlation $c_j$ for the $j$th variable. To facilitate biplot interpretation, it is therefore useful to represent the zero correlation value on the biplot vector~\citep{Graffel40}. More broadly, biplot vectors can be enhanced with a fully calibrated scale with tick marks and tick mark labels to read off the (approximated) correlation. Details on biplot calibration are given by Gower and Hand~\citeyearp{Gower4},~\cite{Graffel17} and \cite{Gower6}. In order to avoid overcrowding the plot, a good solution is to calibrate a biplot by marking off equally spaced correlations (-1.0,-0.9,\ldots,0,\ldots,+0.9,1.0; or finer sub-scales) by dots on the biplot vectors~\citep{Graffel42}.\\

Canonical variates can be interpreted as regression coefficients, obtained by a weighted regression of the biplot axes onto the standardised data (See Graffelman~\citeyearp{Graffel16}, Eqns. (16) and (17)). After convergence of the AGLS algorithm, these regression results can be used to obtain adjusted canonical $X$ and $Y$ variates, denoted by $\bU_a$ and $\bV_a$
respectively:

\begin{equation}
  \bU_{a} = \bXs \bRxx^{-1} \bFs ({\bFs}' {\bRxx}^{-1} \bFs)^{-1} \quad \mbox{ and } \quad \bV_{a} = \bYs \bRyy^{-1} \bGs ({\bGs}' {\bRyy}^{-1} \bGs)^{-1},
  \label{eq:11}
\end{equation}

where $\bXs$ and $\bYs$ refer to the two standardised data matrices, and $\bFs$ and $\bGs$ to standard biplot row and column coordinates respectively. Consequently, adjusted canonical correlations and adjusted canonical coefficients can also be computed. The first adjusted canonical correlation is typically lower than the first canonical correlation obtained by a standard CCA. The regression result~\pref{eq:11} is useful, because it allows the representation of the rows of the original data matrix in the biplots obtained by the adjusted analysis
(see the example in subsection~\ref{subsec:03} below).\\

\section{Examples}
\label{sec:03}

We show several examples that show the improved fit to $\bRxy$ in comparison with a classic canonical analysis using publicly available datasets from social science, geochemistry and medical science.

\subsection{Psychological variables and academic performance}

The data set on psychological measures and academic achievements of 600 college freshmen is a classic example data set for
multivariate analysis, and for canonical correlation analysis in particular. The standard canonical analysis of this data is available
online at \href{https://stats.oarc.ucla.edu}{stats.oarc.ucla.edu} for different statistical environments. The data consists of three psychological variables: locus of control ({\it locus}),
self concept ({\it self}) and {\it motivation}; four academic variables: {\it read}, {\it write}, {\it math}, {\it science} and the demographic variables: {\it female}. The four academic variables are standardised test scores, whereas the last demographic variable is an indicator (1=female,0=male). We first calculate loss and RMSE for CCA and the four different adjustments of the $3 \times 5$ between-set correlation matrix (see Table~\ref{tab:01}) obtained by using the proposed algorithm. A standard canonical analysis has a RMSE of 0.0269. Adjusting the five columns gives minimal RMSE ($<0.00005$), but the column effects are small and most are of similar order of magnitude. For simplicity, we therefore choose to use a $\delta$ adjustment only, which already reduces the RMSE of CCA by a factor of five.  

\begin{table}[ht]
\centering
\begin{tabular}{lccc}
  \hline
Method & $\upsigma$ & RMSE-GLS & RMSE-OLS\\ 
  \hline
CCA          & 0.0108 & 0.0269 & 0.0188 \\ 
CCA-$\delta$ & 0.0004 & 0.0052 & 0.0030 \\ 
CCA-$r$      & 0.0004 & 0.0052 & 0.0030 \\ 
CCA-$c$      & 0.0000 & 0.0000 & 0.0000 \\ 
CCA-$rc$     & 0.0000 & 0.0000 & 0.0000 \\ 
   \hline
\end{tabular}
\caption{Loss ($\upsigma$) and RMSE for approximating the between-set correlation matrix of the psychology-achievement data by canonical correlation analysis and by using scalar, row and/or column adjustments for a rank two approximation.}
\label{tab:01}
\end{table}

\begin{figure}[htb]
  \centering
  \includegraphics[width=0.85\textwidth]{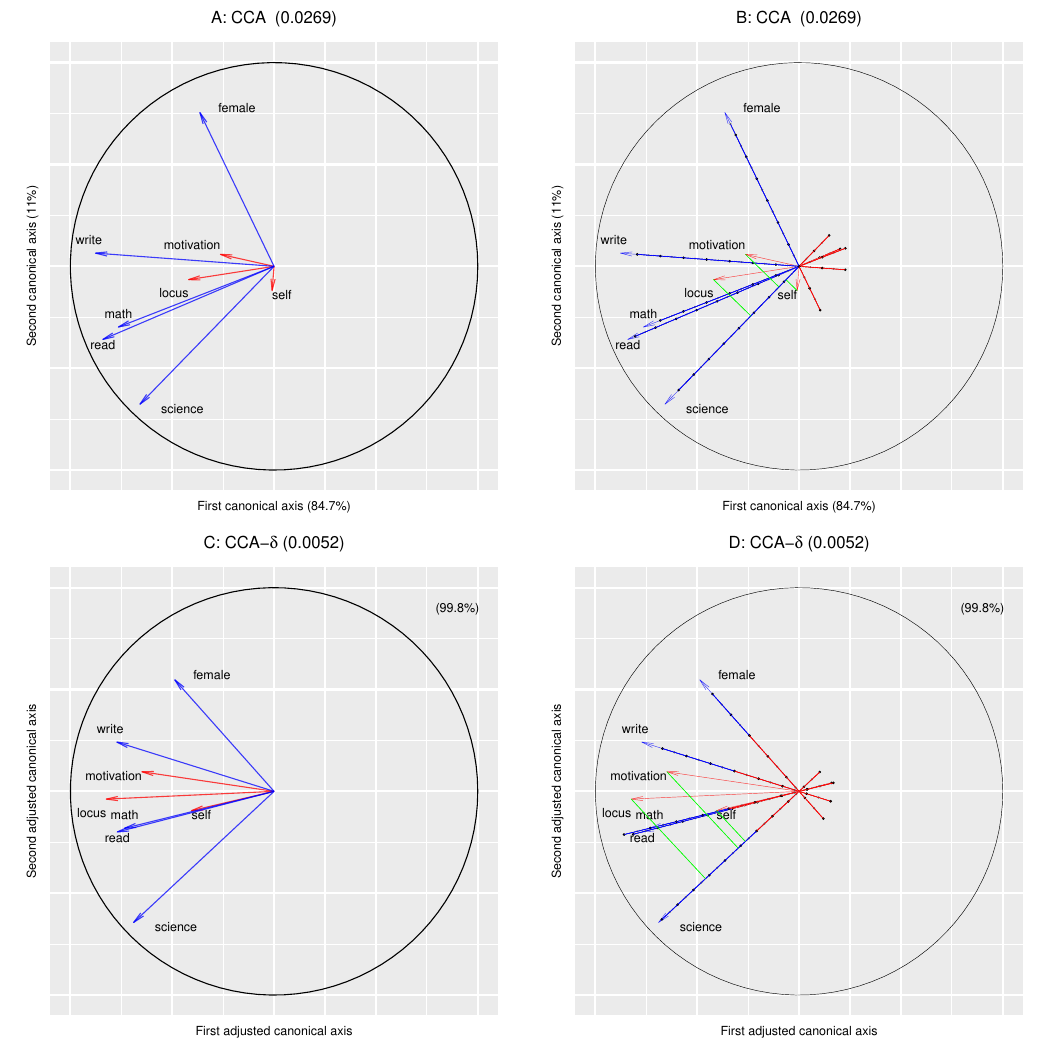}
  \caption{Biplots of the between-set correlation matrix of the psychology-achievement data obtained by CCA (A and B) and CCA-$\delta$ (C and D). Biplot vectors for rows in red, for columns in blue. In panels B and D correlation increments of 0.01, 0.05 and 0.1 scale are marked with grey, dark-grey and (larger) black dots. The negative part of the correlation scale is in red, the positive part in blue. Panels B and D illustrate the approximation of the correlations of the psychological variables with {\it science} with green perpendiculars.}
  \label{fig:01}
\end{figure}

The three canonical correlations are 0.4641, 0.1675 and 0.1040. All three correlations differ significantly from zero as judged by a
permutation test ($p$-values $<$1e-05, 5e-05 and 5.4e-04 respectively). Figure~\ref{fig:01}A shows the biplot of the between-set correlation matrix obtained by classic correlation-based CCA with scaling option $\alpha = 1$. 
\clearpage
The biplot reveals the psychological variables {\it locus} and {\it motivation} correlate negatively with the first canonical variate (i.e., the horizontal axis) and positively with several academic
variables such as {\it read, write} and {\it math}. Figure~\ref{fig:01}B shows the same biplot, but with biplot vectors that are over-plotted with dots showing correlation increments of 0.01 (grey),
0.05 (dark-grey) and 0.10 (black). The dotting of the vectors is convenient for it allows the approximate recovery of the correlations from the plot, while avoiding cluttering it up too
much with fully calibrated scales with tick marks and tick mark labels. In CCA, the origin of the plot represents correlation zero for all variables, and this implies that orthogonal vectors
have, up to the exactness of the approximation, zero correlation. Figure~\ref{fig:01}C shows the biplot obtained by CCA-$\delta$, with estimate $\delta = -0.27$. In this plot all
variables tend to concentrate more around the negative part of the horizontal axis, and the biplot vectors of the psychological variables appear stretched. The origin now represents a correlation of $-0.27$, and in this case it is especially convenient to calibrate the vectors to facilitate the interpretation of the plot as in Figure~\ref{fig:01}D. The sample correlations of {\it self} with the academic achievements are close to zero, and its largest correlation (in the absolute sense) with {\it female} is negative (-0.13). This is clear in the standard CCA biplot in Figure~\ref{fig:01}A, but not in the adjusted plot in Figure~\ref{fig:01}C, unless the correlation scale is added as in Figure~\ref{fig:01}D. The RMSE obtained by CCA-$\delta$ is 0.0052, leading to errors in the approximation of the correlations that are all below 0.01, such that from a numerical and practical point of view, Figure~\ref{fig:01}D is an almost perfect representation of the between-set correlation structure.  

\clearpage

\subsection{Crude oils from sandstone}
\label{subsec:02}

We use chemical data of crude oils from sandstone~\citep{Gerrild} described by Johnson and Wichern~\citeyearp{Johnson2} in the context of linear
discriminant analysis. This data consists of oil samples from three stratigraphic units (Wilhem, Sub-Mulinia and Upper sandstone) using
five components (i.e., $X$ variables): Vanadium ($V$), Iron ($Fe$), Beryllium ($Be$), saturated hydrocarbon ($SH$) and aromatic hydrocarbon ($AH$). These variables were transformed as described by Johnson and Wichern~\citeyearp{Johnson2}, using $\sqrt{Fe}$, $\sqrt{Be}$, and $1/SH$. We code the stratigraphic units with three indicator variables (i.e., $Y$ variables) and study the $5 \times 3$ between-set correlation matrix between chemical components and stratigraphic unit. Table~\ref{tab:02} shows loss and RMSE statistics for the canonical correlation analysis and all adjustments obtained
by the alternating generalised least squares algorithm for rank one and rank two approximations to the between-set correlation matrix.
\begin{table}[ht]  
\centering
\begin{tabular}{lccc|ccc}
  \hline
Method & \multicolumn{3}{c}{Rank 1} & \multicolumn{3}{c}{Rank 2}\\
 & $\upsigma$ & RMSE-GLS & RMSE-OLS & $\upsigma$ & RMSE-GLS & RMSE-OLS\\
  \hline
CCA          & 0.3587 & 0.1546 & 0.1359 & 0.0000 & 0.0000 & 0.0000 \\  
CCA-$\delta$ & 0.1212 & 0.0899 & 3.3484 & 0.0000 & 0.0000 & 0.0000 \\ 
CCA-$r$      & 0.0000 & 0.0000 & 2.1553 & 0.0000 & 0.0000 & 0.0000 \\ 
CCA-$c$      & 0.1212 & 0.0899 & 0.0856 & 0.0000 & 0.0000 & 0.0000 \\ 
CCA-$rc$     & 0.0000 & 0.0000 & 1.8183 & 0.0000 & 0.0000 & 0.0000 \\ 
   \hline
\end{tabular}
\caption{Loss ($\upsigma$) and RMSE according to both a GLS and OLS criterion, for approximating the between-set correlation matrix of Sandstone oils by ordinary and adjusted forms of a canonical correlation analysis. Both rank one and a rank two approximations are considered.}
\label{tab:02}
\end{table}

\begin{figure}[htb]
  \centering
  \includegraphics[width=0.85\textwidth]{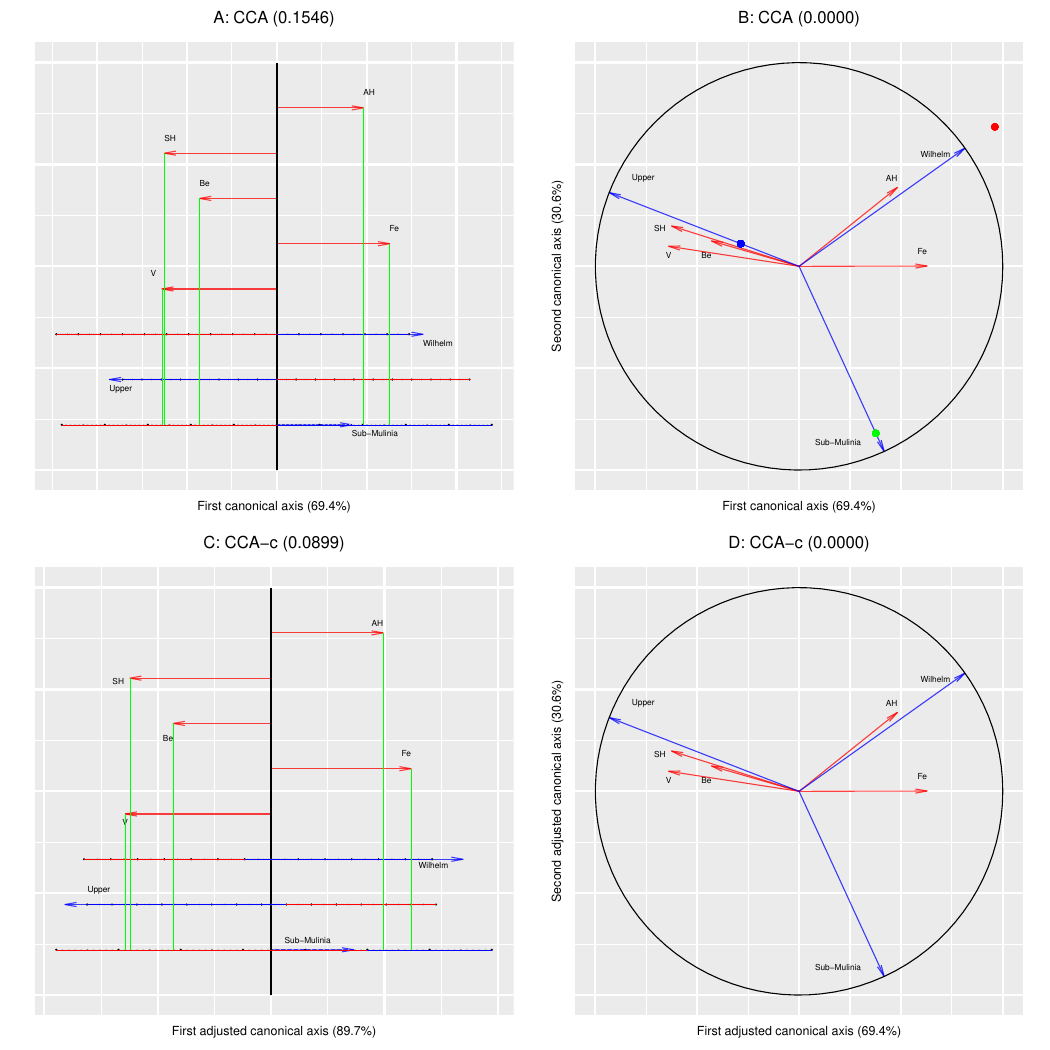}
  \caption{Biplots of the between-set correlation matrix of the sandstone oil data. A: one-dimensional biplot obtained by CCA. B: two-dimensional biplot obtained by CCA. Three dots represent the scores of the canonical $Y$ variates (blue = Upper, red = Wilhelm, green = Sub-Mulinia). C: one-dimensional biplot obtained by CCA-$c$. D: two-dimensional biplot obtained by CCA-$c$. Biplot vectors for rows in red, for columns in blue. The RMSE is given between parentheses in the title of each panel. Sandstone vectors are calibrated with dots. Black (larger), dark-grey and light-grey dots represent increments of 0.1, 0.05 and 0.01 in the correlation scale. Green perpendiculars illustrate the approximation of the between-set correlations.}
  \label{fig:02}
\end{figure}

First note that the covariance matrix of the stratigraphic units indicators is structurally singular and of rank two. This is conveniently solved by using a generalised inverse, the Moore-Penrose inverse~\citep{Muller,Searle}. The between-set correlation matrix is also of rank two and is therefore perfectly fitted by CCA in two dimensions, as witnessed in Table~\ref{tab:02} by zero loss and zero RMSE. In two dimensions, their is no point in performing any adjustment, because the fit obtained by CCA is already perfect. The standard two-dimensional CCA solution is shown in Figure~\ref{fig:02}B. As shown in Figure~\ref{fig:02}D, in this case a two-dimensional CCA-$c$ solution is identical to the CCA solution, and estimated adjustment parameters are just zero. 

\clearpage
Figure~\ref{fig:02}B reveals Upper sandstone oils are characterised by higher values of $V, Be$ and $1/SH$; Wilhelm sandstone oils by higher values of $Fe$ and $AH$; Sub-mulinia sandstone oils are low on $AH$. These and similar conclusions drawn from the biplot are easily verified by making some stratified boxplots, and can also be visualised by over-plotting the biplot with the canonical $Y$ variates, as shown in the figure. The first two canonical correlations are
0.9018 and 0.5989 whereas the third and last canonical correlation is structurally zero. The first two canonical correlations are highly significant as assessed by a permutation test ($p$-values <1e-05). This warrants the use of two dimensions. However, if a one-dimensional approximation is considered, the standard CCA has a large RMSE of 0.1546. Using a scalar adjustment or an adjustment of the columns, the RMSE decreases to 0.0899. Adjusting the rows essentially decreases the RMSE to almost zero (RMSE < 0.00005). However, because the adjustment parameters have an unrestricted range, large values, either negative of positive, can take the correlations outside the (-1,1) interval. This is in fact signalled by calculating the RMSE-OLS criterion. When a column adjustment is used the RMSE's obtained by GLS and OLS criteria are similar, indicating the correlation are not out of range. We therefore choose CCA-$c$ as the best option for a rank one solution.
The corresponding one-dimensional biplots for CCA and CCA-$c$ are shown in Figures~\ref{fig:02}A and ~\ref{fig:02}C, where the biplot vectors have been separated in the second dimension. In this case, the column adjustment allows for an improved representation of rank one of the between-set correlation matrix. Standard CCA (Figure 1A) restricts all biplot vectors to have common origin of zero. The obtained column
adjustments are -0.15, 0.06 and 0.10 for Sub-Mulinia, Upper and Wilhelm sandstones respectively. These adjustments allow for extra flexibility leading to a better fit of the between-set correlation matrix. To facilitate the interpretation of the biplots, the biplot vectors are calibrated, marking off the reference correlation of zero as well as equally spaced increments by dots on the vector, and colouring the positive part of the correlation scale in blue and the negative in red.

\subsection{Cardiovascular disease risk data}
\label{subsec:03}

\begin{figure}[htb]
  \centering
  \includegraphics[width=0.85\textwidth]{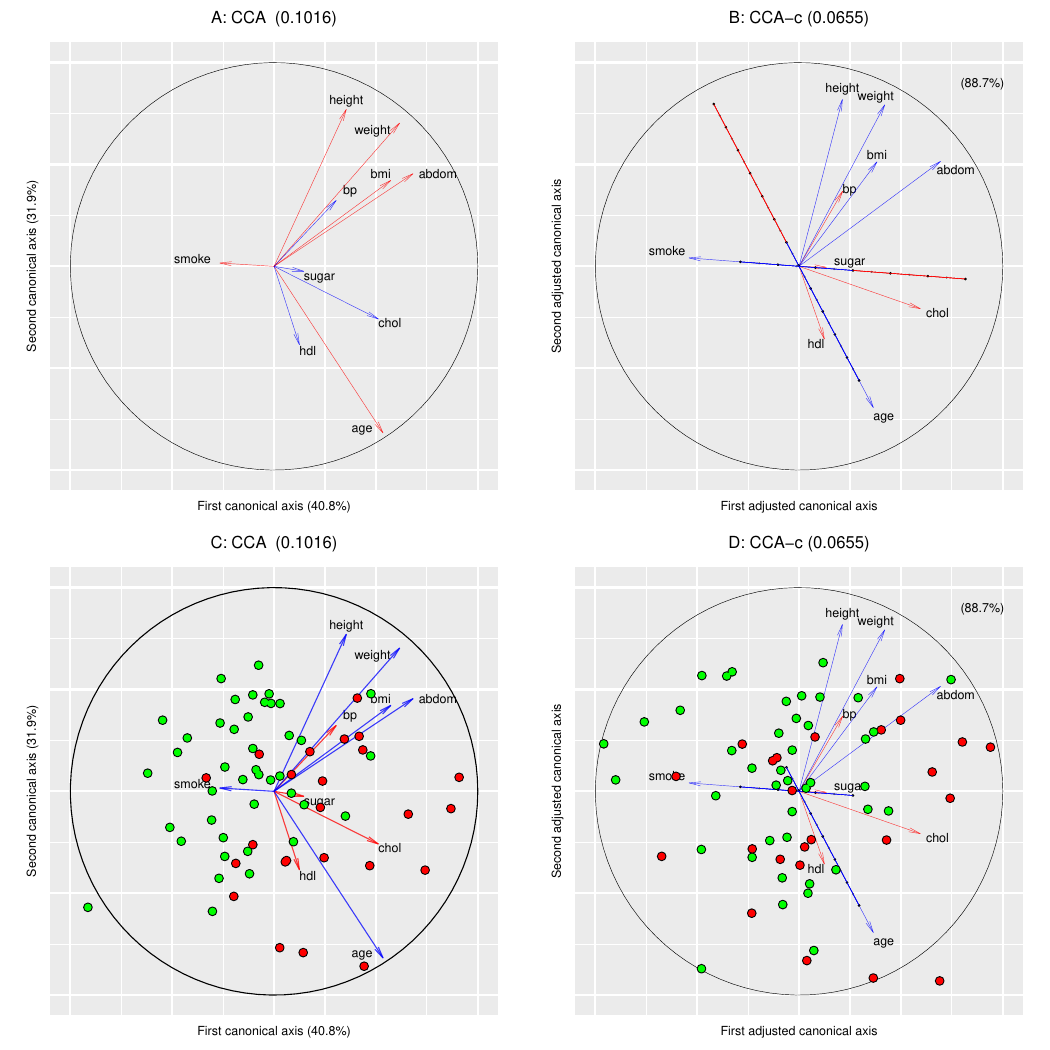}
  \caption{Two-dimensional biplots of the between-set correlation matrix of the cardiovascular dataset. A: standard CCA biplot. B: CCA-$c$ biplot with calibration of {\it age} and {\it smoke}. C: standard CCA biplot with over-plotted canonical variates (individuals). D: CCA-$c$ biplot with over-plotted adjusted canonical variates. Biplot vectors for rows in red, for columns in blue. Low-risk individuals coloured in green, intermediate/high-risk individuals coloured in red. The RMSE is given between parentheses in the title of each panel.}
  \label{fig:03}
\end{figure}

We use the cardiovascular disease risk data of Ferreira et al.~\citeyearp{Ferreira}, consisting of cardiovascular disease related measurements for 71 individuals. We only use the first instance male measurements and distinguish two groups of variables, the four blood-level variables systolic blood pressure ({\it bp} in mmHg), total cholesterol ({\it chol} in mg/dl), high-density lipoprotein ({\it hdl} in mg/dl) and fasting blood sugar ({\it sugar}) and six individual-level variables {\it age}, {\it weight} (kg), {\it height} (m), body mass index ({\it bmi}), abdominal circumference ({\it abdom} in cm), smoking ({\it smoke}) and a health professional cardiovascular risk assessment {\it cvdrisk}. We will use {\it cvdrisk} as a supplementary variable. Figure~\ref{fig:03}A shows the biplot of the between-set correlation matrix obtained by a standard CCA. In two dimensions the goodness-of-fit of this matrix is 72.7\% and the RMSE of the representation is 0.1016. The first two canonical correlations are $0.6082$ and $0.5383$. All four canonical correlations are significant ($p$-values < 0.01) when tested in a permutation test. 
\clearpage
The standard canonical analysis reveals the physical variables {\it weight}, {\it height}, {\it bmi}, and {\it abdom} correlate positively with {\it bp}, whereas {\it age} correlates positively with {\it chol} and {\it hdl}. 

\begin{table}[ht]
\centering
\begin{tabular}{lccc}
  \hline
Method & $\upsigma$ & RMSE-GLS & RMSE-OLS\\
  \hline
CCA          & 0.2480 & 0.1016 & 0.0827 \\ 
CCA-$\delta$ & 0.1974 & 0.0907 & 0.0768 \\ 
CCA-$r$      & 0.1948 & 0.0901 & 0.0774 \\ 
CCA-$c$      & 0.1029 & 0.0655 & 0.0656 \\ 
CCA-$rc$     & 0.1004 & 0.0647 & 0.0660 \\ 
   \hline
\end{tabular}
\caption{Loss ($\upsigma$) and RMSE according to both a GLS and OLS criterion, for approximating the between-set correlation matrix of the cardiovascular dataset by ordinary and adjusted forms of a canonical correlation analysis.}
\label{tab:03}
\end{table}

Values of the loss function and the RMSE for the various types of adjusted canonical analysis are shown in Table~\ref{tab:03}. This table shows that an adjustment of the columns is most interesting. The column adjustments obtained for this model are {\it age = 0.104, weight = 0.045, height = 0.016, bmi = 0.062, abdom = -0.041} and {\it smoke = 0.144}. The biplot of the column-adjusted analysis is shown in Figure~\ref{fig:03}B, where the two variables with the largest adjustments, {\it age} and {\it smoke} have been calibrated to facilitate interpretation. Figure~\ref{fig:03}B shows, for example, that the standard CCA biplot suggests the correlation between {\it smoke} and {\it sugar} to be negative, whereas the calibrated CCA-$c$ biplot correctly reveals this correlation is actually positive. The overall configuration of the biplots in Figures~\ref{fig:03}A and~\ref{fig:03}B is fairly similar, except that the zero correlation point of the biplot vectors is shifted.  On the whole, the adjusted analysis modestly improves the representation of the between-set correlation structure. Figures~\ref{fig:03}C and~\ref{fig:03}D show the same biplots where the individuals have been added by over-plotting the scaled canonical variates according to Eq.~\pref{eq:11}. Individuals assessed as intermediate or high-risk (red points) tend to cluster on the right side of the plot, with higher values for most variables. Canonical variates were scaled by a single constant such as to fit the plot.   

\section{Software}
\label{sec:04}

The algorithm presented in Section~\ref{sec:02} above has been implemented in the function {\tt FitRxy} included in version 1.1.3 of the R-package Correlplot~\citep{GraffelCP}. Function {\tt FitAllModelsRxy} fits all five models for $\bRxy$ and compares their loss and RMSE to assist the choice of which adjustment is most convenient for the data under study. Functions {\tt ggbplot} and {\tt ggtally} can be used to make biplots with calibrated scales. All these functions are documented and illustrated in the vignette of the package.

\section{Discussion}
\label{sec:05}
This article shows that the approximation of the between-set correlation matrix by CCA can be improved by adjusting that matrix by a scalar or by row and/or column effects. The proposed alternating generalised least squares algorithm is essentially an iterated canonical analysis. The fundamental SVD of the canonical analysis (Eq.~\pref{eq:02}) is applied iteratively to the adjusted form of the between-set correlation matrix (Eq.~\pref{eq:05}). The axes extracted in the adjusted analysis are, unlike the standard canonical analysis, not nested; i.e., the sum-of-squares explained by the first dimension of a rank one approximation is different from that explained by the first dimension of a rank two approximation (e.g., see Figures~\ref{fig:02}C
and~\ref{fig:02}D). In standard canonical analysis, in the case of rank two approximation, the total (weighted) sum-of-squares of the between-set correlation matrix can be decomposed into
contributions of the first and the second canonical axis, such that the goodness-of-fit is broken down into independent contributions of each dimension. In the adjusted analysis, only
the goodness-of-fit of the two-dimensional solution can be obtained, but it cannot be broken down in two independent contributions of the two dimensions (e.g., see Figures~\ref{fig:01}and~\ref{fig:03}).\\

In applications of canonical correlation analysis, rank two approximation are commonly used, as these can be plotted in two dimensions. Whenever one of the two sets of variables has just two variables ($p = 2$ or $q=2$, or both), or is of rank two after centring (such as a data matrix with three indicator variables (see Subsection~\ref{subsec:03}) or a three-part composition), the adjustments suggested in the article produce no improvement in the two-dimensional approximation of $\bRxy$, for in that case $\bRxy$ is at most of rank two and can already be presented perfectly in two dimensions with zero RMSE in a standard canonical analysis. The adjustments are therefore useful in the case $\min{(p,q)} > 2$ or for a one-dimensional approximation when $\min{(p,q)} = 2$.\\

The RMSE (or loss) is expected to decrease as more adjustment parameters are used. One may therefore expect the largest reduction when both rows and columns are adjusted, though this option is unattractive for the aforementioned difficulties with the visualisation. When the between-set matrix is not square, choosing to adjust the larger set has more parameters but this does not guarantee that a larger reduction in loss will be obtained than when adjusting the smaller set. Whether it is more beneficial to adjust rows or columns ultimately depends on the data. Adjusting rows or columns should give a loss equal to or smaller than using a single scalar $\delta$. If the order of decrease of the RMSE across the various adjustments is not the expected one, setting a sharper threshold for convergence (smaller $\epsilon$) may fix this.\\

Biplot calibration proves useful for gauging the order of magnitude of the correlations in a biplot and their rate of change across the biplot. The rate of change of the correlations is slower along the shorter biplot vectors. This is explained by the fact the calibration factor (i.e., the vector length that presents a unit change in the scale of interest) depends on the reciprocal of the squared biplot vector length~\citep{Graffel17}. Calibration proves to be particularly important for understanding the biplot of an adjusted analysis, with vectors that no longer have correlation zero at the origin of the plot.\\

Singularity of one or both of the within-set correlation matrices has often hampered the application of canonical correlation analysis. Different scenarios can give rise to singularity and it is convenient to distinguish these. If one set of the variables consists of indicator variables that represent all categories of a categorical variable (such as a geological stratum in the second example), the rank of the corresponding within-set correlation matrix goes down by 1 due to the single linear constraint that the variables sum to one. Likewise, in a compositional setting, data vectors arise that sum to a constant, typically fractions summing to one. Compositional data, when transformed by the centred logratio transformation~\citep{Aitchison3}, also has a singular covariance matrix whose rank is also reduced by 1. Both these cases can be conveniently solved by using a generalised inverse such as the Moore-Penrose inverse; the problem is not the high dimensionality (many more variables than observations) but the existence of a structural linear relationship between the variables. The structural singularity due to high dimensionality ($max(p,q) \gg n$) has been addressed by regularization or by replacing the within-set correlation matrices with identity matrices.

\section{Disclosure Statement}

The author reports there are no competing interests to declare.

\section{Data availability statement}

The three datasets used in this article are publicly available. The data set on psychological measures and academic achievements is available at \href{https://stats.oarc.ucla.edu}{stats.oarc.ucla.edu} and also included in the R package Correlplot 1.1.3. The sandstone oil data is listed in Johnson and Wichern~\citeyearp{Johnson2}. The cardiovascular disease risk data is available at Mendeley Data \href{https://dx.doi.org/10.17632/vhgyn5yk4g.1}{10.17632/vhgyn5yk4g.1}. 

\bibliographystyle{Chicago}

\bibliography{BetweenSetArxiv}

\end{document}